\documentclass{appolb}
\usepackage{epsfig}

\newcommand{\be}{\begin{equation}}
\newcommand{\ee}{\end{equation}}
\newcommand{\bea}{\begin{eqnarray}}
\newcommand{\eea}{\end{eqnarray}}
\newcommand{\ba}{\begin{array}}
\newcommand{\ea}{\end{array}}
\newcommand{\beann}{ \begin{eqnarray*}}
\newcommand{\eeann}{ \end{eqnarray*}  }
\newcommand{\eq}[1]{Eq.~(\ref{#1})}
\newcommand{\fig}[1]{Fig.~\ref{#1}}
\def\tr{{\rm Tr}}
\newcommand{\lag}[1]{{\mathcal{L}}_{\rm {#1}}}
\def\mK{m_K}
\def\FK{F_K}
\def\mbar{\overline{m}}
\def\rmA{{\rm A}}

\def\rmP{{\rm P}}
\def\ZP{Z_{\rmP}}
\def\ZA{Z_{\rmA}}
\def\dd{{\rm d}}
\def\GeV{\,{\rm GeV}}
\def\MeV{\,{\rm MeV}}
\def\fm{\,{\rm fm}}
\def\gbar{\bar{g}}
\def\mbar{\overline{m}}
\def\mbars{\mbar_s}
\def\mbaru{\mbar_u}

\def\mref{m_{\rm ref}}
\def\Mref{M_{\rm ref}}
\def\muu{m_u}
\def\Mu{M_u}
\def\md{m_d}
\def\Md{M_d}
\def\ms{m_s}
\def\Ms{M_s}
\def\MSb{\overline{{\rm MS}}}

\def\Lmax{L_{\rm max}}

\def\Nf{N_{\rm f}}

\begin{document}
\eqsec  
\title{{\vskip -50pt
\mbox{} \hfill CERN-PH-TH/2005-214 \\
\mbox{} \hfill HU-EP-05/72 \\
\mbox{} \hfill SFB/CPP-05-74 }
\vskip 25pt
Lattice computation of the strange quark mass in QCD
\thanks{Presented at
XXIX International Conference of Theoretical Physics, \emph{Matter To The Deepest: Recent Developments In Physics of Fundamental Interactions},
Ustron, 8--14 September 2005, Poland}
}
\author{Francesco Knechtli
\address{CERN, Physics Department, TH Division, CH-1211 Geneva 23, Switzerland}
}
\maketitle
\begin{abstract}
We present a determination of the strange quark mass using lattice QCD. Particular
focus is put on the definition and renormalization of the mass. The latter is done
non-perturbatively, using a recursive finite-size scaling technique.
The hadronic regime of QCD, where the kaon mass is used as input of the calculation,
is connected with the perturbative regime, where the strange quark mass can be
translated into the $\MSb$ scheme. A summary plot of the present lattice computations
using dynamical (sea) quarks is included.
\end{abstract}

\section{Introduction}

Quantum chromodynamics (QCD) as the theory of strong interactions between the
$\Nf=3$ light quark flavours $q=u,d,s$
(we treat the $c,b,t$ quarks as infinitely heavy)
is described through the Lagrange density
\bea
 \lag{QCD}(g_0,m_q) & = & -\frac{1}{2g_0^2}\tr\{F_{\mu\nu}F_{\mu\nu}\} +
 \sum_{q} \bar{q}\,(\gamma_\mu(\partial_\mu + A_\mu) + m_q)\,q \,. \label{Lqcd}
\eea
The bare gauge coupling $g_0$ and bare quark masses $m_q$ are the free parameters of QCD.
In order to make predictions, we need
after a suitable regularization, e.g. on a Euclidean space--time lattice,
to fix the free parameters through a set of $\Nf+1$ physical quantities
from experiment. For example, we can take as experimental input hadronic quantities
such as
\beann
\overbrace{
\left[\ba{c} {F_\pi}     \\
             {m_\pi}     \\
             {\mK}       \ea\right]}^{\mbox{Experiment}} \; &
\ba{c}\lag{QCD}(g_0,m_q)\\ \Longleftrightarrow\\
      \ea\; &
\overbrace{
\left[\ba{c} {\Lambda_{\rm QCD}}   \\
             {\hat{M}=(\Mu+\Md)/2} \\
             {\Ms}        \ea\right]}^{\mbox{QCD parameters (RGI)}} \;+\;
\overbrace{
\left[\ba{c} {m_{\rm Nucleon}}     \\
             {B_K}           \\
             {\ldots} \ea\right]}^{\mbox{Predictions}}
\eeann
and, using \eq{Lqcd}, extract from the high-energy regime of QCD the
so-called renormalization group invariant (RGI) parameters associated with the
running renormalized gauge coupling and masses. In addition a number of other
predictions can be made for phenomenologically relevant quantities. 
A crucial point becomes immediately clear: we need a tool to relate the hadronic,
low-energy regime of QCD with the high-energy ($>2$--$10\GeV$) regime, where perturbation
theory in the gauge coupling applies. In the following we will demonstrate that
such a tool is provided by the Euclidean space--time lattice.
%
\begin{figure}[t]
\centerline{\epsfig{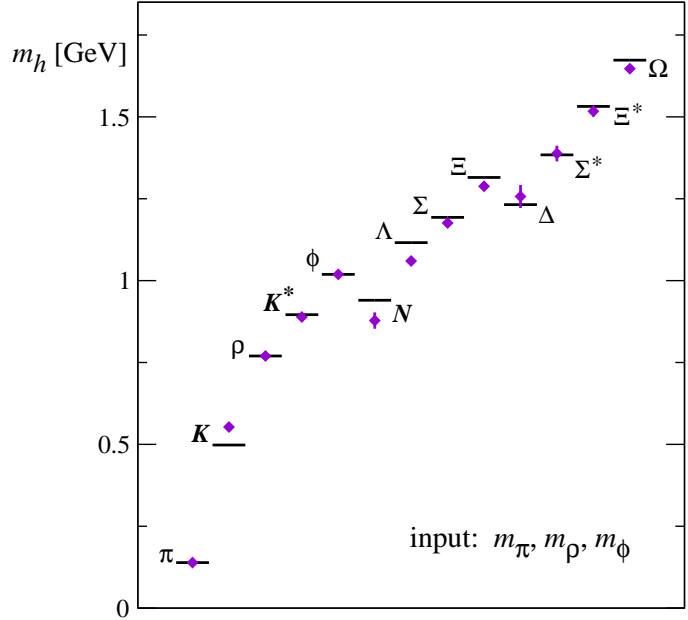}}
\caption{The hadron spectrum from quenched lattice QCD.
\label{f_had_masses}}
\end{figure}
%

The predictive power of lattice QCD can be seen in \fig{f_had_masses}, which is
taken from \cite{MLPLOT}. The data points represent various hadron masses $m_h$
computed by the CP-PACS Collaboration \cite{CPPACSSPEC}.
The input used in the lattice simulations
are $m_\rho$, $m_\pi$ and $m_\phi$, which fix the lattice spacing $a$,
the mass of the degenerate up and down quarks, and the strange quark mass
(a chiral extrapolation is needed for the mass of the up and down quarks).
The hadron spectrum is extrapolated to the continuum limit $a\to0$.
Given that quark polarization effects have been neglected
(the so-called quenched approximation),
the agreement between data points and experimental values marked by the horizontal lines
is remarkable.

In this article we will be concerned with the computation of the strange quark mass
from lattice QCD. In the limit of vanishing quark masses (chiral limit)
QCD possesses a large chiral symmetry. This symmetry is spontaneously broken and
the spectrum contains eight pseudo-scalar Goldstone bosons, which correspond to the
observed eight lightest hadrons ($\pi$'s, $K$'s, $\eta$). The latter are not exactly
massless due to the non-zero quark masses. Quark masses are the explicit symmetry-breaking
parameters \cite{GL1,HLCHIPT} and they are treated as perturbations of the
chiral limit in the framework of chiral perturbation theory \cite{GL2}. At lowest
order the quark mass ratios $\muu/\md=0.56$ and $\ms/\md=20.1$
can be determined from the pion and kaon masses \cite{GMOR,WMR}. Chiral perturbation
theory, though, cannot determine the absolute scale of the quark masses;
lattice computations
are required for this. The strange quark mass can be determined almost directly through
lattice computations; few assumptions are needed, which we will explain below.
For the up and down quark masses, more difficult chiral extrapolations are needed.

The value of the strange quark mass is required as an input for
phenomenological predictions of the Standard Model,
such as the CP-violating ratio $\epsilon^\prime/\epsilon$.
Also for physics beyond the Standard Model the quark masses
are very important parameters.

\section{The strange quark mass from lattice QCD}

In the following we will use Wilson's formulation of lattice QCD, including
Symanzik's O($a$) improvement. A review of principles of lattice computations
can be found in Ref. \cite{MLAQCD}. The Wilson formulation explicitly breaks the
chiral symmetry through an O($a$) term in the Wilson--Dirac operator \cite{BMMRT}. As a
consequence the quark mass receives an additive renormalization. However, it is well
understood how the chiral Ward identities can be implemented on the lattice up to
cut-off effects, which are O($a^2$) in the improved theory; for a short review see
\cite{RHPHD}.

A renormalized quark mass on the lattice,
translated into the continuum $\MSb$ scheme, can be defined through
\bea
 \mbar^{\MSb}(\mu) & = & Z_m(g_0,a\mu)(m-m_{\rm crit}) \,.
\eea
A definition that avoids the determination of the additive mass renormalization $m_{\rm crit}$
is through the partial conservation of the axial currents (PCAC) relation, e.g. in the form
\bea
 \partial_\mu(\bar{u}\gamma_\mu\gamma_5s)_{\rm lat} & = & 
 (\muu+\ms)_{\rm lat}(\bar{u}\gamma_5s)_{\rm lat} + {\rm O}(a^2) \,, \label{pcac}
\eea
where $A_\mu=(\bar{u}\gamma_\mu\gamma_5s)_{\rm lat}$ is an improved axial current
\cite{CA},
$P=(\bar{u}\gamma_5s)_{\rm lat}$ a pseudo-scalar density, and
$m_q^{\rm lat}$ are the bare lattice current quark masses.
A renormalized quark mass is defined from the PCAC relation \eq{pcac}
using the renormalized axial current $(A_{\rm R})_\mu=\ZA A_\mu$ and
pseudo-scalar density $P_{\rm R}=\ZP P$ and comparing the bare and renormalized
relations:
\bea
 (\mbaru+\mbars)^{\MSb}(\mu) & = & (\muu+\ms)_{\rm lat}\,
 \frac{\ZA(g_0)}{\ZP^{\MSb}(g_0,a\mu)} \,. \label{mbar}
\eea
The renormalization factor $\ZA(g_0)$ has been calculated non-perturbatively, using the
chiral Ward identities \cite{ZA}.
In order to compute $\ZP^{\MSb}(g_0,a\mu)$, we could apply
perturbation theory. At one loop,
\bea
 \ZP^{\MSb}(g_0,a\mu) & = & 1+\frac{g_0^2}{4\pi}[(2/\pi)\ln(a\mu)+k] +
                            {\rm O}(g_0^4) \,, \label{zp1loop}
\eea
where $k$ is a calculable constant that depends on the lattice regularization and
the renormalization scheme employed.
Bare perturbation theory is known to be unreliable at the coupling accessible to
simulations $g_0\simeq1$ and the systematic error of the perturbative expansion is
difficult to assess. A non-perturbative determination of $\ZP$ avoids these problems.

The running of the renormalized gauge coupling $\gbar$ and quark masses $\{\mbar_q\}$
depends on the renormalization scheme employed and is controlled by the
renormalization group equations (RGEs)
\bea
\mu\frac{\dd\gbar}{\dd\mu}=\beta(\gbar)\,, &&
\mu\frac{\dd\mbar_q}{\dd\mu}=\tau(\gbar)\mbar_q \,,
\eea
in terms of the $\beta$ and $\tau$ functions. They are in general non-perturbatively
defined. Their perturbative expansions are
\bea
 \beta(\gbar) &
 _{\mbox{$\stackrel{\displaystyle\sim}{\scriptstyle \gbar\rightarrow0}$}} &
 -\gbar^3\{b_0+b_1\gbar^2+b_2\gbar^4+...\} \,,\\
 \tau(\gbar)  &
 _{\mbox{$\stackrel{\displaystyle\sim}{\scriptstyle \gbar\rightarrow0}$}} &
 -\gbar^2\{d_0+d_1\gbar^2+...\} \,.
\eea
If the renormalization conditions are imposed at zero quark mass \cite{WEIN}
the $\beta$ and $\tau$ functions do not depend on the mass. One such scheme,
the $\MSb$ scheme, is the one most commonly used for perturbative QCD computations
and there the $\beta$ and $\tau$ functions are known up to the 4-loop coefficients 
$b_3$ \cite{MSbeta1,MSbeta2} and $d_3$ \cite{MStau1,MStau2}.
The RGI quark masses are defined through
\bea
 M_q = \mbar_q(\mu)\left(2b_0\gbar^2(\mu)\right)^{-d_0/(2b_0)}
 \exp\left\{-\int_0^{\gbar(\mu)}\dd x\left[
 \frac{\tau(x)}{\beta(x)}-\frac{d_0}{b_0x}\right]\right\} \,.
\eea
Together with the $\Lambda$ parameter, they
are independent of the renormalization scale $\mu$ and, in fact, any physical
quantity, whose total dependence on $\mu$ vanishes, can be considered as a 
function of $\Lambda$ and $\{M_q\}$. The RGI masses are the same
in all massless renormalization schemes.

A straightforward implementation of the definition \eq{mbar} on the lattice poses a
scale problem. In order to compute $\ZP$,
the matrix element $\langle 0 | P | K^+ \rangle$ is needed and for this
the spatial lattice size $L$ has to be large enough with respect to a typical scale like the
kaon decay constant $\FK$, so as to avoid finite-volume effects.
The renormalization scale $\mu$ has to be large enough to make
contact with perturbation theory and at the same time $\mu$ has to be small with respect to
the cut-off $1/a$ in order to avoid large cut-off effects. We get a set of inequalities
\cite{RNPR}
\beann
 L \quad \gg \hspace{1.5em}
 \frac{1}{\FK}\sim\frac{1}{0.2\GeV} \hspace{1.0em}\gg \quad
 \frac{1}{\mu} &\sim& \frac{1}{10\GeV} \quad
 \gg a \,,
\eeann
which imply $L/a \gg 50$. But lattices that big cannot be simulated.
One elegant solution to this problem is to take a finite-size effect as the physical
observable, that is to identify $L=1/\mu$, with the only requirement $L/a\gg1$.
The renormalization scale $\mu$ is changed recursively in steps by factors of 2.
Individually at each step the continuum limit can be taken.
One therefore speaks of the recursive finite-size scaling (or step scaling)
technique \cite{LWW}.
We take the strange quark mass as an example: a running mass
$\mbars(\mu)$ is defined\footnote{
For a specific renormalization scheme implementing this idea, see below.}
that runs with the system's size $\mu=1/L$. Starting at a reference maximal value $\Lmax$,
the system's size $L$ is halved $n$ times until contact with the
perturbative regime is made; the RGI mass $\Ms$ can then be extracted:
      \begin{eqnarray*}
        \Lmax={\rm const.}/\FK = {\rm O}(0.5\fm): \qquad
        \longrightarrow \quad
        &\Lmax\,\mbars(\mu=1/\Lmax)& \quad
                \\
                &\downarrow&   \\
      &\Lmax\,\mbars(\mu=2/\Lmax)&   \\[-1ex]
      &\downarrow&   \\[-1ex]
      \mbox{always } a\mu=a/L\ll 1     &\vdots&   \\[-1ex]
      &\downarrow&   \\
      &\Lmax\,\mbars(\mu=2^n/\Lmax)& \\
      \mbox{perturbation theory}\;\;\; &\downarrow&   \\
      &\Lmax\,\Ms =  \#&
    \end{eqnarray*}
If the reference scale $\Lmax$ is expressed in units of the kaon decay constant $\FK$,
the result is a number for $\Ms/\FK$, which is scheme-independent.
%
\begin{figure}[t]
\vspace{-2.0cm}
\centerline{\epsfig{file=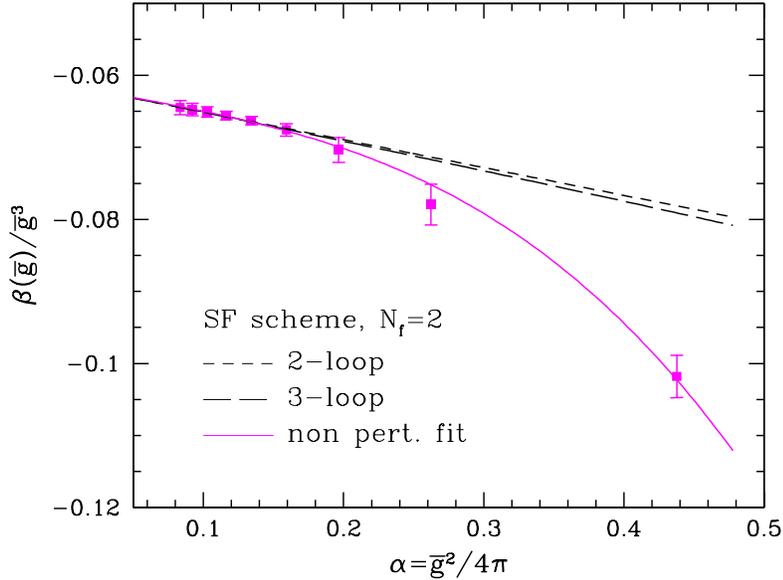,width=11cm}}
\vspace{-0.7cm}
\caption{The non-perturbative $\beta$ function in the SF scheme
for $\Nf=2$ flavours of massless quarks.
\label{f_beta}}
\end{figure}
%

A particularly suitable renormalization scheme to implement the finite-size
scaling technique is the Schr\"odinger functional (SF). It has been defined
for QCD in Refs. \cite{LNWW,SS} and is the ``work horse'' of the ALPHA Collaboration.
QCD is formulated in a box where Dirichlet boundary conditions are imposed on the
fields at Euclidean time $x_0=0$ and $x_0=T$. They provide an infrared cut-off
proportional to $1/T$ to the frequency spectrum of quarks and gluons, making it
possible to perform simulations, and hence impose renormalization conditions,
at zero quark mass.
A renormalized gauge coupling $\gbar(L)$, running with the scale $\mu=1/L$,
is defined through a variation of the
effective action with respect to a change of the boundary gluon fields
\cite{ALPHAQUENCH,ALPHA}. Keeping $\gbar(L)$ fixed, while changing the bare
lattice parameters, is equivalent to keeping the physical size $L$ (and
hence the renormalization scale $\mu$) fixed. 

The recursive finite-size scaling technique in the SF has been applied to compute
the running of the renormalized coupling with $\Nf=2$ flavours of massless quarks
\cite{ALPHA}. The non-perturbative $\beta$ function is shown by the points in
\fig{f_beta}, together
with a non-perturbative fit. It deviates from the 3-loop $\beta$ function for
$\alpha_{\rm SF}(L)=\gbar^2(L)/(4\pi)>0.25$.
%
\begin{figure}[t]
\vspace{-2.0cm}
\centerline{\epsfig{file=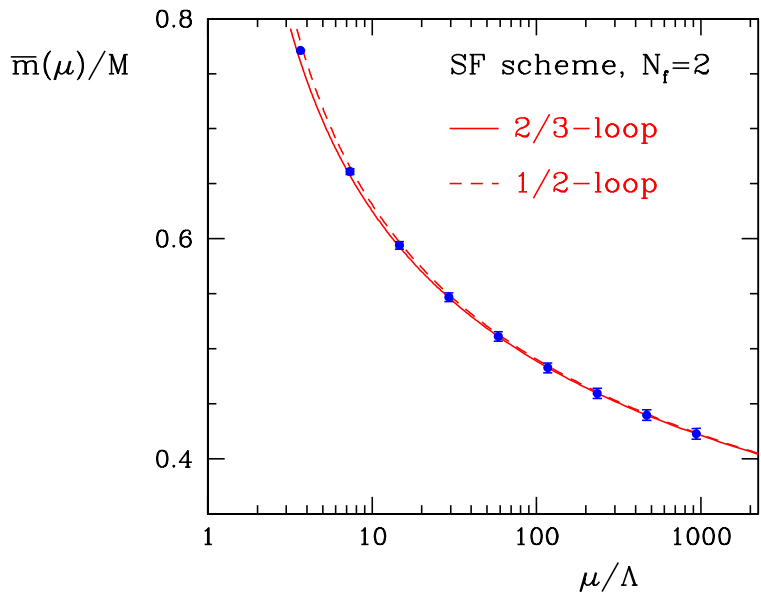,width=12cm}}
\vspace{-1.7cm}
\caption{The running quark mass in the SF scheme
for $\Nf=2$ flavours of massless quarks. Plot taken from \cite{MASS}.
\label{f_mrun}}
\end{figure}
%

The renormalization factor $\ZP$ can be defined and computed in the SF 
scheme \cite{ZP,SWZP,CAP}.
The running of the renormalized quark mass is extracted from it through
\bea
 \frac{\mbar^{\rm SF}(\mu)}{\mbar^{\rm SF}(\mu/2)} & = &
 \left.\lim_{a\to0}
 \frac{\ZP^{\rm SF}(g_0,2L/a)}{\ZP^{\rm SF}(g_0,L/a)}\right|_{\mu=1/L} \,,
\eea
which is independent of the quark flavour $q$. The results for $\Nf=2$ flavours
of massless quarks have been presented in detail in Ref. \cite{MASS} and are
shown by the points in \fig{f_mrun}.
The running starts at the low energy $\mu=1/\Lmax\sim0.4\GeV$ corresponding
to the leftmost point and extends to high energies, where perturbation theory
can be safely applied to extract the RGI mass.
The curves are obtained by using the perturbative expressions
for the $\tau/\beta$ functions at the loop order indicated in the legend. 
In this case perturbation theory
works surprisingly well down to small energies.
For the running gauge coupling, instead, at $\alpha_{\rm SF}(\Lmax)=0.367$,
clear deviations between non-perturbative and perturbative curves
can be seen in \fig{f_beta}.

In the following we summarize the computation of the strange quark mass
done in Ref. \cite{MASS}.
The steps involved are visible in the equation
\bea
\mbars^{\MSb}(\mu_2) = \left(\frac{\mbar(\mu_2)}{M}\right)_{\rm PT}^{\MSb}
\times
\left(\frac{M}{\mbar(\mu_1)}\right)_{\rm NP}^{\rm SF}
\times
(\ms)_{\rm lat}\frac{\ZA(g_0)}{\ZP(g_0,a\mu_1)} \,. \label{defmbar}
\eea
The scale in the $\MSb$ scheme is conventionally $\mu_2=2\GeV$.
The first factor on the right-hand side of \eq{defmbar}
is the perturbative (PT) evolution in the $\MSb$ scheme.
The second factor is
the non-perturbative (NP) running in the SF scheme that starts at the
hadronic scale $\mu_1=1/\Lmax$. The third factor is the renormalized (at the scale $\mu_1$)
strange quark mass in the SF scheme and it has to be computed
at several lattice spacings to take the continuum limit. This part
involves in principle
\begin{enumerate}
\item
simulations of $u$, $d$ and $s$ quark flavours at their physical masses, and
\item
setting the overall scale of the simulations, i.e.
determining the lattice spacing $a$ in $\MeV^{-1}$.
\end{enumerate}

Because of computer cost and technical difficulties, simulations of two light and
non-degenerate $u$ and $d$ quarks with an additional $s$ quark have not been possible so far.
There are very promising developments reported in \cite{ML05}, which
show that these difficulties can be overcome in the near future.
For the time being, the strategy of \cite{MSQUENCH} is adopted. Simulations
are performed with two degenerate flavours at a reference mass $\mref$.
The latter is tuned so that the pseudo-scalar (PS), made of two identical flavours,
has a mass
\bea
 m_{\rm PS}(\mref,\mref) & = & \mK = 
 \frac{1}{2} \left( m_{K^+}^2+m_{K^0}^2\right)_{\rm QCD} = 495\MeV\,.
 \label{defmref}
\eea
The subscript ``QCD'' means that an estimate of the electromagnetic effects
has been subtracted from the experimental numbers \cite{MSQUENCH} in order
to obtain a pure QCD kaon mass, as we have on the lattice.

The lattice spacing could be set by computing on the lattice the kaon decay
constant, a number $a\FK$, and dividing it by the experimental value of $\FK$,
which can be obtained from the decay rate of $K^+\longrightarrow\mu^+\nu_\mu$.
Instead, in the present computation,
the lattice spacing is set through the scale $r_0$ extracted from the
static quark potential \cite{R0}. Data for $r_0/a$ as well as pseudo-scalar
masses needed to tune $\mref$ in \eq{defmref} are taken from \cite{QCDSF},
where they are available at three lattice spacings in the approximate range
$0.092$--$0.071\fm$.
The data for $r_0/a$ are extrapolated to zero quark mass \cite{ALPHA}. The
phenomenological value is $r_0=0.5\fm$, obtained from potential models.

The results for the reference RGI quark mass show a strong
dependence on the lattice spacing \cite{MASS}. No systematic continuum extrapolation
is therefore performed, instead a continuum estimate is given:
\bea
 \Mref & = & 72(3)(13)\MeV \,, \label{Mreference}
\eea
where the value and the first error are the result at the smallest lattice spacing,
the second error is a systematic one and is the difference with the value
at the largest lattice spacing. The latter dominates. The result \eq{Mreference}
is consistent with the one obtained in the quenched ($\Nf=0$) theory\footnote{
Also for the $\Lambda$ parameter there is no significant difference between
$\Nf=0$ and $\Nf=2$ \cite{ALPHA}.}
\cite{MSQUENCH}. It is assumed to hold in the $\Nf=3$ theory as well.
In order to translate \eq{Mreference} into a value for the strange quark mass,
the Gell-Mann--Oakes--Renner formula is used \cite{GMOR,GL1,GL2}:
\bea
\mK^2 \,=\, \left(\hat{M}+\Ms\right)\,B_{\rm RGI} & = &
2\Mref\,B_{\rm RGI} \,, \label{mK}
\eea
where $\hat{M}=\frac{1}{2}(\Mu+\Md)$ and $B_{\rm RGI}$ is a constant of the
chiral Lagrangian. Together with $\Ms/\hat{M}=24.4(1.5)$ \cite{HL},
\eq{mK} can be solved for $\Ms$, and using 4-loop running in the $\MSb$ scheme
yields
\bea
 \mbars^{\MSb}(\mu=2\GeV) & = & 97(22)\MeV \,. \label{mstrange}
\eea
The $\mbars^{\MSb}$
values for the three lattice spacings are displayed in \fig{f_summary} (filled
red squares).

\section{Conclusions}

%
\begin{figure}[t]
\centerline{\epsfig{file=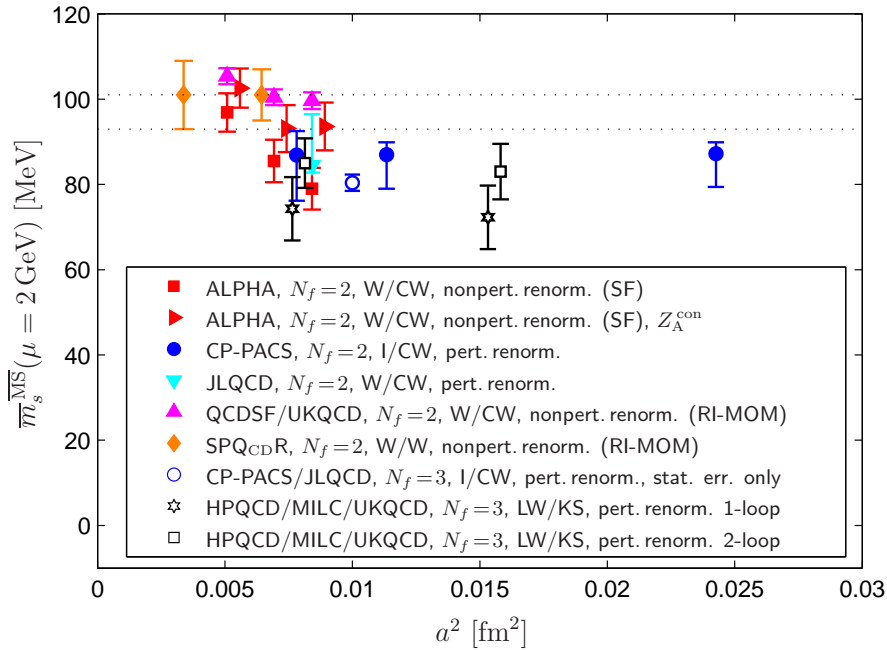,width=12cm}}
\caption{Summary plot of the strange quark mass from dynamical simulations
with $\Nf=2$ and $\Nf=3$. \label{f_summary}}
\end{figure}
%

In \fig{f_summary} we show a summary of the lattice determinations of the
strange quark mass with dynamical (sea) quarks at finite values of the lattice
spacing $a$. This is an update of the plot of Refs. \cite{MASS,MASS2}.
There are points obtained with $\Nf=2$ dynamical quarks
\cite{MASS,QCDSF,CPPACS,JLQCD,SPQCDR} and $\Nf=2+1$ \cite{CPPJL3,MILC}.
In the legend we indicate
the gauge action/fermion action employed and
whether non-perturbative or perturbative renormalization has
been used.
The dictionary for the gauge actions is:
W: Wilson action; I: Iwasaki action;
LW: 1-loop tadpole improved L{\"u}scher--Weisz action.
The dictionary for the fermion actions is:
W: Wilson action; CW: Wilson-clover action; KS: Asqtad staggered action.
Except the Wilson fermion action used in \cite{SPQCDR}, all the other lattice
formulations have O($a^2$) discretization errors.
The non-perturbative renormalization is done in the SF or the
RI-MOM \cite{RIMOM} scheme.
The physical kaon mass is always used as input to fix the strange quark mass,

There is a systematic effect due to the use of perturbative renormalization,
which leads to values of the quark mass smaller than those for non-perturbative
renormalization. Figure \ref{f_summary} also shows the impact in the quark mass values of
\cite{MILC} when the perturbative renormalization factor $\ZP$ in \eq{zp1loop}
is evaluated at two loops \cite{ZP2l}: the mass values increase by 15\%.
There is good agreement between data from different non-perturbative 
renormalization procedures
and, in fact, between them and the $\Nf=0$ result \cite{MSQUENCH}
marked by the dotted lines.
We also mention the most recent compilation of the strange quark mass from QCD
sum rules \cite{NARISON}, which quotes $\mbars(2\GeV)=99(28)\MeV$, in agreement
with the range of the lattice results; see also \cite{MJ1,MJ2,BCK}.

In conclusion, \fig{f_summary} demonstrates that non-perturbative renormalization
is essential for a reliable lattice determination of the strange quark mass and that
data at smaller lattice spacing(s) are needed for a systematic continuum extrapolation.

{\bf Acknowledgements} I am very grateful to Ulli Wolff for his comments on the
manuscript. I also express many thanks to the Organizers for their warm hospitality
and the very pleasant conference.


\begin{thebibliography}{99}

\bibitem{MLPLOT}
M. L{\"u}scher,
\emph{Annales Henri Poincar{\'e}} {\bf 4}, 197 (2003) [{\tt hep-ph/0211220}].

\bibitem{CPPACSSPEC}
CP-PACS Collaboration, S. Aoki et al.,
\emph{Phys. Rev.} {\bf D67}, 034503 (2003) [{\tt hep-lat/0206009}].

\bibitem{GL1}
J. Gasser and H. Leutwyler,
\emph{Phys. Rep.} {\bf 87}, 77 (1982).

\bibitem{HLCHIPT}
H. Leutwyler,
Principles of chiral perturbation theory [{\tt hep-ph/9406283}],
Lectures given at the Hadrons 94 Workshop, Gramado, Brazil, 1994.

\bibitem{GL2}
J. Gasser and H. Leutwyler,
\emph{Nucl. Phys.} {\bf B250}, 465 (1985).

\bibitem{GMOR}
M. Gell-Mann, R.J. Oakes, B. Renner,
\emph{Phys. Rev.} {\bf 175}, 2195 (1968).

\bibitem{WMR}
S. Weinberg,
\emph{Trans. N.Y. Acad. Sci.} {\bf 38}, 185 (1977).

\bibitem{MLAQCD}
M. L{\"u}scher,
Advanced Lattice QCD [{\tt hep-lat/9802029}],
Lectures given at Les Houches Summer School in Theoretical Physics, Les Houches, France, 1997.

\bibitem{BMMRT}
M. Bochicchio, L. Maiani, G. Martinelli, G.C. Rossi, M. Testa,
\emph{Nucl. Phys.} {\bf B262}, 331 (1985).

\bibitem{RHPHD}
R. Hoffmann,
Chiral properties of dynamical Wilson fermions, Ph.D. thesis
[{\tt hep-lat/0510119}].

\bibitem{CA}
M. Della Morte, R. Hoffmann, R. Sommer,
\emph{JHEP} {\bf 03}, 029 (2005) [{\tt hep-lat/0503003}].

\bibitem{ZA}
ALPHA Collaboration, M. Della Morte, R. Hoffmann, F. Knechtli, R. Sommer, U. Wolff,
\emph{JHEP} {\bf 07}, 007 (2005) [{\tt hep-lat/0505026}].

\bibitem{WEIN}
S. Weinberg,
\emph{Phys. Rev.} {\bf D8}, 3497 (1973).

\bibitem{MSbeta1}
T. van Ritbergen, J.A.M. Vermaseren, S.A. Larin,
\emph{Phys. Lett.} {\bf B400}, 379 (1997) [{\tt hep-ph/9701390}].

\bibitem{MSbeta2}
M. Czakon,
\emph{Nucl. Phys.} {\bf B710}, 485 (2005) [{\tt hep-ph/0411261}].

\bibitem{MStau1}
K.G. Chetyrkin,
\emph{Phys. Lett.} {\bf B404}, 161 (1997) [{\tt hep-ph/9703278}].

\bibitem{MStau2}
J.A.M. Vermaseren, S.A. Larin, T. van Ritbergen,
\emph{Phys. Lett.} {\bf B405}, 327 (1997) [{\tt hep-ph/9703284}].

\bibitem{RNPR}
R. Sommer,
Non-perturbative renormalization of QCD [{\tt hep-ph/9711243}],
Lectures given at 36th Internationale Universit{\"a}tswochen f{\"u}r Kernphysik und Teilchenphysik, Schladming, Austria, 1997.

\bibitem{LWW}
M. L{\"u}scher, P. Weisz, U. Wolff,
\emph{Nucl. Phys.} {\bf B359}, 221 (1991).

\bibitem{LNWW}
M. L{\"u}scher, R. Narayanan, P. Weisz, U. Wolff,
\emph{Nucl. Phys.} {\bf B384}, 168 (1992) [{\tt hep-lat/9207009}].

\bibitem{SS}
S. Sint,
\emph{Nucl. Phys.} {\bf B421}, 135 (1994) [{\tt hep-lat/9312079}].

\bibitem{ALPHAQUENCH}
M. L{\"u}scher, R. Sommer, P. Weisz, U. Wolff,
\emph{Nucl. Phys.} {\bf B413}, 481 (1994) [{\tt hep-lat/9309005}].

\bibitem{ALPHA}
ALPHA Collaboration, M. Della Morte, R. Frezzotti, J. Heitger, J. Rolf,\\ R. Sommer, U. Wolff,
\emph{Nucl. Phys.} {\bf B713}, 378 (2005) [{\tt hep-lat/0411025}].

\bibitem{ZP}
K. Jansen, C. Liu, M. L{\"u}scher, H. Simma, S. Sint, R. Sommer, P. Weisz,\\ U. Wolff,
\emph{Phys. Lett.} {\bf B372}, 275 (1996) [{\tt hep-lat/9512009}].

\bibitem{SWZP}
ALPHA Collaboration, S. Sint and P. Weisz,
\emph{Nucl. Phys.} {\bf B545}, 529 (1999) [{\tt hep-lat/9808013}].

\bibitem{CAP}
ALPHA Collaboration, S. Capitani, M. L{\"u}scher, R. Sommer, H. Wittig,
\emph{Nucl. Phys.} {\bf B544}, 669 (1999) [{\tt hep-lat/9810063}].

\bibitem{MASS}
ALPHA Collaboration, M. Della Morte, R. Hoffmann, F. Knechtli,\\ J. Rolf, R. Sommer, I. Wetzorke,
U. Wolff,
\emph{Nucl. Phys.} {\bf B729}, 117 (2005) [{\tt hep-lat/0507035}].

\bibitem{ML05}
M. L{\"u}scher,
Plenary talk at 23rd
International Symposium on Lattice Field Theory: Lattice 2005, Trinity College, Dublin, Ireland,
[{\tt hep-lat/0509152}].

\bibitem{MSQUENCH}
ALPHA Collaboration, J. Garden, J. Heitger, R. Sommer, H. Wittig,
\emph{Nucl. Phys.} {\bf B571} 237 (2000) [{\tt hep-lat/9906013}].

\bibitem{R0}
R. Sommer,
\emph{Nucl. Phys.} {\bf B411}, 839 (1994) [{\tt hep-lat/9310022}].

\bibitem{QCDSF}
QCDSF and UKQCD Collaborations, M. G{\"o}ckeler, R. Horsley, A.C. Irving, D. Pleiter, P.E.L. Rakow, G. Schierholz, H. St{\"u}ben,
[{\tt hep-ph/0409312}].

\bibitem{HL}
H. Leutwyler,
\emph{Phys. Lett.} {\bf B378}, 313 (1996) [{\tt hep-ph/9602366}].

\bibitem{MASS2}
ALPHA Collaboration, M. Della Morte, R. Hoffmann, F. Knechtli, J. Rolf,\\ R. Sommer, I. Wetzorke, U. Wolff,
[{\tt hep-lat/0509073}].

\bibitem{CPPACS}
CP-PACS Collaboration, A. Ali Khan et al.,
\emph{Phys. Rev.} {\bf D65}, 054505 (2002) [{\tt hep-lat/0105015}].

\bibitem{JLQCD}
JLQCD Collaboration, S. Aoki et al.,
\emph{Phys. Rev.} {\bf D68}, 054502 (2003) [{\tt hep-lat/0212039}].

\bibitem{SPQCDR}
D. Be{\'c}irevi{\'c}, B. Blossier, Ph. Boucaud, V. Gim{\'e}nez, V. Lubicz, F. Mescia,\\ S. Simula, C. Tarantino,
[{\tt hep-lat/0510014}].

\bibitem{CPPJL3}
CP-PACS and JLQCD Collaborations, T. Ishikawa et al.,
\emph{Nucl. Phys. Proc. Suppl.} {\bf 140}, 225 (2005) [{\tt hep-lat/0409124}].

\bibitem{MILC}
HPQCD and MILC and UKQCD Collaborations, C. Aubin et al.,
\emph{Phys. Rev} {\bf D70}, 031504 (2004) [{\tt hep-lat/0405022}].

\bibitem{RIMOM}
G. Martinelli, C. Pittori, C.T. Sachrajda, M. Testa, A. Vladikas,
\emph{Nucl. Phys.} {\bf B445}, 81 (1995) [{\tt hep-lat/9411010}].

\bibitem{ZP2l}
HPQCD Collaboration, Q. Mason, H. Trottier, R. Horgan,
Plenary talk at 23rd
International Symposium on Lattice Field Theory: Lattice 2005, Trinity College, Dublin, Ireland,
[{\tt hep-lat/0510053}].

\bibitem{NARISON}
S. Narison,
[{\tt hep-ph/0510108}].

\bibitem{MJ1}
M. Jamin, J.A. Oller, A. Pich,
\emph{Eur. Phys. J.} {\bf C24}, 237 (2002) [{\tt hep-ph/0110194}].

\bibitem{MJ2}
E. Gamiz, M. Jamin, A. Pich, J. Prades, F. Schwab,
\emph{Phys. Rev. Lett.} {\bf 94}, 011803 (2005) [{\tt hep-ph/0408044}].

\bibitem{BCK}
 P.A. Baikov, K.G. Chetyrkin, J.H. K{\"u}hn,
\emph{Phys. Rev. Lett.} {\bf 95}, 012003 (2005) [{\tt hep-ph/0412350}].

\end{thebibliography}
\end{document}